\documentclass[newstyle,twocolumn,proceedings]{rmaa}
\usepackage{rmaacite}
\renewcommand{\P}[1]{%
\ifnum#1=1\hbox{OW~168--326E}\fi
\ifnum#1=2\hbox{OW~167--317}\fi
\ifnum#1=3\hbox{OW~163--317}\fi
\ifnum#1=5\hbox{OW~158--323}\fi
\ifnum#1=0\hbox{OW~171--334}\fi}
\makeatletter
\title{Adaptive Optics imaging of Quasar Hosts\altaffilmark{1}}
\author{I. M\'arquez\affil{Instituto de Astrof\'{\i}sica de Andaluc\'{\i}a 
(CSIC), Granada (Spain)} 
and Patrick Petitjean\affil{Institut d'Astrophysique de Paris (France)}}
\altaffiltext{1}{Based on data obtained at the Canada-France-Hawaii telescope
which is operated by CNRS of France, NRC of Canada and the University of
Hawaii.}
\fulladdresses{
\item I. M\'arquez: Instituto de Astrof\'{\i}sica de
Andaluc\'{\i}a (CSIC), Apdo 3004, 18080 Granada (Spain) (isabel@iaa.es).
\item P. Petitjean: Institut d'Astrophysique de Paris (CNRS), 98 bis Bd Arago, F-75014 Paris (France) and UA CNRS 173 -- DAEC, Observatoire de Paris-Meudon, F-92195 Meudon
Cedex (France) (petitjean@iap.fr)}
\shortauthor{M\'arquez \& Petitjean}
\shorttitle{AO imaging of QSO Hosts}

\abstract{We present the results of adaptive-optics imaging of 12 ($z$~$<$~0.6) 
quasars in the H and K bands using the PUEO system mounted on the 
Canada-France-Hawaii telescope. The QSOs ($m_{\rm V}$~$>$~15.0) themselves are 
used as reference for the correction. The images, obtained under poor seeing 
conditions, have typical spatial resolution of $FWHM$~$\sim$~0.3~arcsec before 
deconvolution. The deconvolved H-band image of PG~1700$+$514 has a spatial resolution
of 0.16~arcsec and reveals a wealth of details on the companion and
the host-galaxy. The comparison between the HST and CFHT images of PG~1700+514
shows how powerful and competitive ground-based AO can be. 
Close companions and obvious signs of interactions are found in 4 out of the 
12 objects. The 2D images of 3 of the host-galaxies unambiguously reveal bars 
and spiral arms. The morphology of the other host-galaxies are difficult to determine 
from 1D surface brightness profiles and deeper images are needed that could be 
obtained with AO systems on 10~m class telescopes.\\
Analysis of mocked data shows that elliptical galaxies are always
recognized as such, whereas disk hosts can be missed for small disk scale 
lengths and large QSO contributions.
}

\resumen{Presentamos los resultados de im\'agenes en las bandas H y K con 
\'optica adaptativa (OA) usando el sistema PUEO en el CFHT, de 12 quasares
($z$~$<$~0.6). El quasar mismo ($m_{\rm
V}$~$>$~15.0) es utilizado como referencia para la correcci\'on. Las 
im\'agenes, obtenidas en pobres condiciones de seeing, tienen resoluciones 
espaciales t\'{\i}picas de $FWHM$~$\sim$~0.3 segundos de arco antes de la 
deconvoluci\'on. La imagen deconvolucionada de PG~1700$+$514 tiene una 
resoluci\'on espacial de  0.16 segundos y revela con todo detalle tanto la 
galaxia albergadora como la compa\~nera. La comparaci\'on entre las im\'agenes 
HST y CFHT de PG~1700+514 muetra cu\'an poderosa puede ser la OA, tanto m\'as 
utilizada en telescopios de la clase de 10~m. 4 de los 12 objetos presentan 
compa\~neros y signos obvios de interacci\'on. Las im\'agenes 2D de 3 de 
ellos indican claramente la presencia de barras y brazos espirales. 
La morfolog\'{\i}a del resto es dif\'{\i}cil de determinar a partir del 
perfil de brillo; se necesitan im\'agenes m\'as profundas, que podr\'an obtenerse con OA en telescopios de la clase de 10~m.\\
El an\'alisis de im\'agenes simuladas muestra que galaxias inicialmente 
el\'{\i}pticas siempre se reconocen como tales, mientras que las galaxias de 
disco pueden ser identificadas como el\'{\i}pticas para longitudes de escala 
del disco peque\~nas y para contribuciones importantes del QSO.}
\listofauthors{I.~M\'arquez \& P.~Petitjean}
\indexauthor{M\'arquez, I.}
\indexauthor{Petitjean, P.}
\begin{document}

%% This command is necessary to typeset the title, abstract, etc. 
\maketitle

%%
%% And here starts the text....
%%
\section{Introduction}
\label{sec:intro}
Evidence that nearby bright galaxies contain massive dark objects in
their center has become increasingly compelling over the last few
years and early suggestions that a tight correlation exists between
the mass of the dark object and the mass of the bulge \cite{kr95}
have been convincingly corroborated \cite{magorrian,ferrarese}.  It
is thus possible that AGN activity is a usual episode of the history
of most, if not all, present-day bright galaxies.  One way to
investigate this is to determine the luminosity and morphology of
galaxies hosting quasars. In addition, this gives clues on the range
of conditions needed for strong nuclear activity to occur.

Several studies of quasar host-galaxies have been performed
with the main aim of distinguishing between radio-loud and radio-quiet
quasars, either from IR ground-based observations
\cite{dunlop,mcleod95a,taylor} or by using HST imaging
\cite{disney,bahcall97,mclure99,kirkhakos}.
The detection and analysis of host-galaxies is difficult even from space
\cite{bahcall94,bahcall95,mcleod95b,mcleod00}.  Indeed, the
determination of the PSF and the subtraction of the point source image
are crucial in this work. Differenciation between the two classical
profiles, either an exponential disk or a de Vaucouleur power-law, is
effective only in the regions close to the center, or in the far-wings
of the PSF (see Fig.~1 of \pcite{mclure00}). Because of seeing limitation, 
PSF subtraction is the main limitation in determining host-galaxy morphologies 
from the ground. With the advent of adaptive optics however, it is possible to
alleviate this limitation \cite{stockton98,aretxaga,hutchings99,marquez01}. 
In addition, observing in the infrared minimizes the difference in luminosity 
between the host and nucleus again improving our ability to determine the host 
morphology. PSF determination is still a major problem but the difficulties are
balanced by the prospect of using 10~m class telescopes
which will provide higher sensitivity and better spatial resolution.
\par\noindent 
In this contribution we present the summary of the
results of a pilot programme aiming at testing the capabilities of
adaptive optics in this field. The details can be found in \scite{marquez01}.

\section{Sample selection and data} 
\label{sec:sample}
In order to use adaptive optics correction quasars were selected such
that the nuclei is bright enough to be used as the wavefront
reference point source. The sample of radio-quiet quasars were all PG
quasars with m$_b$ $<$ 16.5 and with redshift less than 0.6. The
radio-loud objects were selected from 3C, 4C, B2 and PKS catalogues
with the same magnitude and $z$ criteria. The final objects observed
(see Table 1 in \pcite{marquez01}) were selected based upon the
suitability for the observing conditions on the observing runs.

We used the CFHT adaptive optics bonnette (PUEO) and the IR camera KIR
on May 1998 and May 1999.  The weather conditions were poor during
both runs and the FWHM of the seeing PSF was never better than 0.8
arcsec. The adaptative-optics correction was performed on the QSOs
themselves.  The final images after the usual process of NIR-imagin
reduction, have a typical resolution of $FWHM$~$\sim$~0.3
arcsec. After each science observation an image of a star with similar
magnitude as the QSO was taken in order to determine the PSF and use
it to deconvolve the images.  Due to rapid variations in the wheather
conditions however, it was not always possible to follow this
predefined procedure.
\par\noindent
A synthetic PSF function, derived from the stellar images was used to
deconvolve each of the images.  As it was not always possible to apply
a standard procedure due to fluctuating seeing conditions, a careful
although, somewhat arbitrary choice of the PSF had to be done.  In 
\scite{marquez01} we have shown that the best PSF to be used for deconvolution 
 is that obtained using the star with the FWHM closest
to that of the science exposure. In general,
this illustrates the crucial role played by a careful PSF
determination in AO observations.

\section{Analysis}
\label{sec:analysis}
In each of the images, we first masked out the companion objects and
the ghosts due to the telescope.  We then obtained the surface
brightness profiles of the galaxies using the IRAF\footnote{IRAF is 
the Image Analysis and Reduction Facility made available to the 
astronomical community by the National Optical Astronomy
Observatories, which are operated by the Association of Universities
for Research in Astronomy (AURA), Inc., under contract with the
U.S. National Science Foundation.} task {\sl ellipse}.  The resulting
profiles were fitted over the radius range from 3 times the FWHM of
the PSF up to the point where the galaxy surface brightness level
falls below 2 $\times \sigma$ of the background level.  We have
systematically fitted an exponential disk and a de Vaucouleurs
r$^{1/4}$ law.  Out of the 10 objects for which we can
extract some morphological information, two of the host-galaxies are
most probably barred spirals, the rest being ellipticals or very early
type spirals.  We note that in general the number of
points we can use to fit either profile is not large enough to
unambiguously distinguish between the two fitted profiles. At small
radii, the excess of light between the observed profile and the model
disk can be due to the presence of a bulge. The morphology 
is determined considering both the 2D
luminosity spatial distribution and the 1D profile. It is apparent
that to discriminate between both morphologies, the S/N ratio must be
high at large radii.

The magnitudes of the hosts have been derived by integrating the
$r^{1/4}$ profiles for all the objects. Indeed, it can be seen
that there is an excess of light at small radius
compared to the disk profile for all objects. This suggests that
in our sample, the disk galaxies have also a strong bulge and/or
a strong bar. This is confirmed by the 2D luminosity distribution.
We have 
also subtracted a scaled version of the most suitable PSF for each
nucleus imposing a non-negative profile in the center (see \pcite{marquez}).
The resulting host magnitudes, computed by
integrating the PSF-subtracted images, are in good agreement
with those obtained from the profile fitting.

The results of the analysis are given in Table~2 in \scite{marquez01}.
We also give the number of objects (probably companions)
found within 5~ and 10~arcsec from the quasar down to $m_{\rm
H}$~=~20.5 and the maximum radial distance (in arcsec and kilo-parsec) to
which the host is detected at a significance level of 3$\sigma$ above
the background. 

We concentrate in the results on PG~1700+514. The description of the
remaining objects can be found in \scite{marquez01}.

\section{PG1700+514}
\label{sec:pg1700}
PG~1700+514 is one of the most infrared-luminous, radio-quiet BAL
quasar \cite{turnshek85,turnshek97}. Ground-based imaging revealed an
extension about 2~arcsec north-east of the quasar \cite{stickel}
which was shown by adaptive-optics imaging and follow-up spectroscopy
to be a companion with a redshift 140~km~s$^{-1}$ blueward of the
quasar \cite{stockton98}. NICMOS observations lead Hines et al. (1999) to
argue that the companion is a collisionally induced ring galaxy.  The
fit to the SED and the Keck spectrum of the companion imply that the
light is emitted by an old population of stars plus a 85 Myr old
star-burst \cite{canalizo}. Note however that the H-band
flux ($m_{\rm H}$~$\sim$~16.6) deduced from HST imaging is much larger
than that predicted by the model.  \scite{stockton98} showed
that the inclusion of embedded dust can produce a spectral-energy
distribution that is consistent with both the optical
spectrophotometry and the IR photometry.

\begin{figure}
\includegraphics[width=\columnwidth]{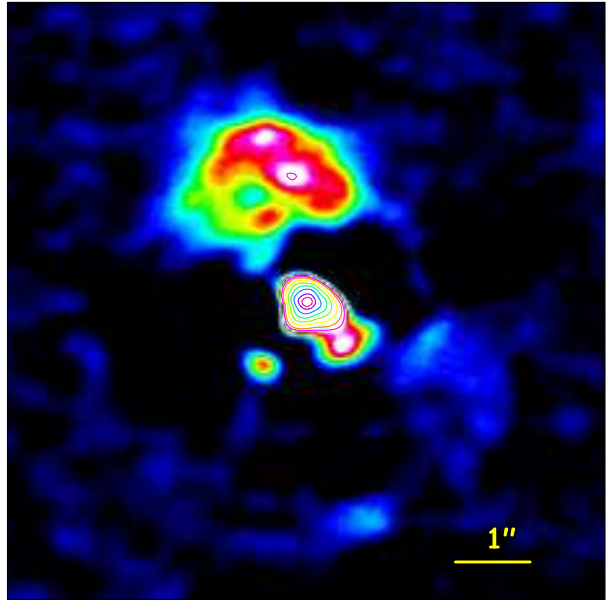}
\caption{Image of PG~1700+514 after deconvolution using
the PSF given by the star with the PSF closest to that of the quasar.
The resulting image has a final 
resolution of $FWHM$~=~0.16~arcsec. 
}
\label{pg1700}
\end{figure}

The image obtained at CFHT is shown in Fig.~\ref{pg1700}. 
We confirm the findings by \scite{stockton98} that the companion has the
appearance of an arc with several condensations.  We used different
PSF to deconvolve the image. The best deconvolution is obtained using
the star with the FWHM closest to that of the AGN (0.30
arcsec). The image has a final resolution of 0.16 arcsec and is
probably the best image obtained yet on this object.  The companion is
seen as a highly disturbed system with a bright nucleus and a
ring-like structure; the nucleus beeing decentered with respect to the
ring. The host-galaxy is clearly seen around the quasar with a bright
extension to the south-west, first noted by \scite{stickel} and 
clearly visible in the optical images by \scite{stockton98}. In addition, 
we detect a bright knot to
the south-east which is not seen in the NICMOS data probably because
of the presence of residuals in the PSF subtraction.  
The comparison between the HST and CFHT images of PG~1700+514
shows how powerful AO can be, and bodes well for the use of the technique 
on 10~m-class telescopes.
No obvious relation is found between the near-IR image and the 
radio map \cite{hutchings92}.

\section{Mock data}
\label{sec:mock}
We have fitted the 1-D profiles of the host-galaxies with either exponential 
or de~Vaucouleurs laws.
In order to test our fitting procedure, we have generated images of
model elliptical and disk galaxies with scale-lengths and effective
surface brightness within the range derived from the data.  The same
orientation and axis ratio is given to all of them. A point source
is added in the center of the galaxy to mimic the quasar.
An appropriate amount of noise is added, and then the images are 
convolved with a typical observed PSF. The mocked images are analyzed
in the same way as real data.

We first note that an elliptical galaxy is always recognized as
such by the fitting procedure, whereas a disk-galaxy
is better fitted by a $r^{1/4}$ law when the unresolved 
point-source contributes more than half the total light.
This means that,
at least with data of similar quality to those presented here, 
the fraction of elliptical galaxies in the sample may be overpredicted. 
Going deeper, at least 0.5 to 1 magnitude, should help solve this problem 
as it is apparent that the distinction between spiral and 
elliptical profiles is easier when the galaxy is detected at larger distances
from the central point-source. 

It is interesting to note that the output magnitudes are brighter than
the input in both cases, elliptical or disk galaxies. The reason for this is
probably the difficulty in determining the extension of the PSF wings
which, if not subtracted properly, will artificially increase the flux
of the host-galaxy. In the case of spirals, the difference is as large
as 0.6 magnitudes when the contribution of the point-source is the
same as the contribution of the host-galaxy. 
For the ellipticals, the difference is less
but still important when the QSO dominates the total flux.

Note that the ratio between the QSO and the host-galaxy luminosities 
is expected to increase with redshift. The above bias tends to imply
that host-galaxy luminosities could be overestimated.

\section{Conclusions}
\label{sec:conclusions}
AO imaging in the H and K bands has been used to study the morphology
of QSO host-galaxies at low and intermediate redshifts ($z$~$<$~0.6).
We detect the host-galaxies in 11 out of 12 quasars (5 radio-quiet and
7 radio-loud). The images, obtained under poor seeing conditions, and
with the QSOs themselves as reference for the correction, have typical
spatial resolution of $FWHM$~$\sim$~0.3~arcsec before
deconvolution. Close companions and obvious signs of interactions are
found in 4 objects. The 2D images of 3 of the host-galaxies
unambiguously reveal bars and spiral arms.  For the other objects, it
is difficult to determine the host-galaxy morphology on the basis of
one dimensional surface brightness fits alone. In the best case, the
deconvolved H-band image of PG~1700$+$514 (with a spatial resolution
of 0.16~arcsec) reveals a wealth of detail on the companion and the
host-galaxy, and is probably the best-quality image of this object
thus far.

We have simulated mocked images of host-galaxies, both spirals and
ellipticals, and applied the same analysis as to the data.  Disk hosts
can be missed for small disk scalelengths and large QSO
contributions. In this case, the host-galaxy can be misidentified as
an elliptical galaxy. Elliptical galaxies are always recognized as
such, but with a luminosity which can be overestimated by up to 0.5
magnitudes.  The reason for this is that the method used here tends to
attribute some of the QSO light to the host. This is also the case for
disk galaxies with a strong contribution of the unresolved component.

\acknowledgements This work is
financed by DGICyT grants PB96-0921, PB98-0521 
and AYA2001-2089. 
Financial support to
develop the present investigation has been obtained through the Junta
de Andaluc\'{\i}a TIC-114. 

%% When using the rmaacite package, the \bibitem command should be of
%% the format: 
%%
%%             \bibitem[AUTHOR<YEAR>]{KEY} 
%%
%% so that the \cite{KEY}, etc. commands will work properly. 
%% 
%% If you are doing the citations manually, then you can just use
%% `\bibitem{}' instead. This will give you a warning about
%% `multiply-defined labels' which you can safely ignore.
%% 

\end{document}